# A biologically inspired computational trust model for open multi-agent systems which is resilient to trustor population changes

**Zoi Lygizou**
Hellenic Open University, Patra, Greece
std084140@ac.eap.gr (Corresponding author)
https://orcid.org/0000-0001-8414-7963

**Dimitris Kalles**
Hellenic Open University, Patra, Greece
kalles@eap.gr
https://orcid.org/0000-0003-0364-5966

**Abstract**

Current trust and reputation models continue to have significant limitations, such as the inability to deal with agents constantly entering or exiting open multi-agent systems (open MAS), as well as continuously changing behaviors. Our study is based on CA, a previously proposed decentralized computational trust model from the trustee's point of view, inspired by synaptic plasticity and the formation of assemblies in the human brain. It is designed to meet the requirements of highly dynamic and open MAS, and its main difference with most conventional trust and reputation models is that the trustor does not select a trustee to delegate a task; instead, the trustee determines whether it is qualified to successfully execute it. We ran a series of simulations to compare CA model to FIRE, a well-established, decentralized trust and reputation model for open MAS, under conditions of continuous trustee and trustor population replacement, as well as continuous change of trustees' abilities to perform tasks. The main finding is that FIRE is superior to changes in the trustee population, whereas CA is resilient to the trustor population changes. When the trustees switch performance profiles FIRE clearly outperforms despite the fact that both models' performances are significantly impacted by this environmental change. Findings lead us to conclude that learning to use the appropriate trust model, according to the dynamic conditions in effect could maximize the trustor's benefits.

1. Introduction

Trust and reputation models have been widely employed in the identification of malicious, untrusted behavior in multi-agent systems (MAS) [1]. As more computational systems have moved towards large-scale, open and dynamic architectures [5], trust models have also been developed to meet those systems' requirements.

However, current trust management techniques continue to have significant limitations, such as the inability to cope with continuously changing behaviors [1]. In open MAS, agents constantly enter and exit the system, a challenging situation for common trust models to deal with [4]. In dynamic environments, agents' behavior can change quickly, and this must be identified by consumer agents relying on trust and reputation models to choose profitable provider agents for their interactions [8]. Note here, that the terms *providers* and

*trustees* refer to the agents who provide services, while the terms *consumers* and *trustors* refer to the agents who use these services.

We base our work on CA [9], a trust model, which views trust from the perspective of the trustee; the trustor does not select a trustee, as in the more conventional trust modeling approach, but it is the trustee which decides whether it is skilled to perform the required task. CA is inspired by biology, specifically the synaptic plasticity in human brain, which allows neurons to form coherent groups called assemblies (which is why it is called "Create Assemblies (CA)". As it is previously discussed in [9], the fact that the trustor is not required to choose confers a few key advantages on CA approach in highly dynamic and open systems. The trustee's calculated trust value is information that the trustee can carry and use in every application it joins, giving it an advantage in dealing with the issue of mobility, an issue still considered an open challenge [6]. Choosing trustees by trustors necessitates extensive trust information exchange among agents, which leads to more communication time [11]. Several agents may be reluctant to share their private judgments about other agents [10], while giving up an agent's privacy about how it values the services of others can have a major impact [5]. Importantly, in CA approach, agents do not exchange information which naturally creates the expectation that CA model is immune to various kinds of disinformation, a common condition in most agent societies.

In this paper, we report a series of simulation experiments, aiming to compare CA's robustness to that of FIRE, under dynamic trustees profiles and continuous agents' entering and exiting of the system. We chose FIRE as a reference model, because it is an established trust and reputation model for open MAS that, like CA model follows the decentralized approach. Furthermore, FIRE is representative of the conventional approach, in which trustors choose trustees (TCT approach), to contrast with CA approach, in which trustees are not chosen by trustors.

Our findings lead us to conclude that CA's main advantage is its resilience to the trustor population change. The rest of the paper is organized as follows. In Section 2, we discuss related background. Section 3 describes the testbed as well as the methodology we used for the experiments conducted. Section 4 presents the experiments' results, section 5 discusses our findings and finally, Section 6 concludes, highlighting potential future work.

## 2. Background
### 2.1. The nature of Open and dynamically changing MAS

Open MAS are usually defined as systems in which agents continuously exit or enter the system [4]. Yet, several researchers [7, 2, 12] have recognized task openness as another type (apart from agent openness) of MAS openness. Specifically in [12], openness is viewed as the system's ability to deal with dynamic changes, like the arrival of new tasks and resources. Under this scope, we adopt the following more general definition. Open MAS are systems that can handle three changes of entities: removal, addition, and evolution [4], where entities can be agents, tasks, or resources.

In our experiments, we considered all three modifications. Specifically for evolution we have let an agent's performance to change over time, as it learns or forgets how to execute tasks. Yet, we have not investigated task and resources openness, leaving it for future work.

### 2.2. The FIRE model

FIRE [5] is a well-known trust and reputation model that follows the distributed, decentralized approach. It incorporates the following four modules:

- Interaction Trust (IT): an agent's trustworthiness is determined by the evaluator based on previous interactions with it.
- Witness Reputation (WR): a target agent's trustworthiness is evaluated based on the opinions of witnesses, i.e. other agents that have interacted with it.
- Role-based Trust (RT): the evaluator assesses trustworthiness based on role based relationships with the target agent and on available domain knowledge, as norms and regulations.
- Certified Reputation (CR): trustworthiness is determined by the evaluator using third-party references stored by the target agent, available on demand.

### 2.3. The CA model

In [9], the authors formally described a novel computational trust model from the trustee's point of view, inspired by synaptic plasticity, a biological process responsible for the formation of assemblies, large population of neurons in human brain, that are thought to imprint cognitive information, such as words and memories.

According to this model, the trustor (i.e. the agent which requests the task to be executed) broadcasts a message containing the following information: a) the type of work to be done (task category), and b) a set of task requirements. When a trustee receives a request message, it establishes and maintains a connection with a weight $w \in [0,1]$ denoting the connection's strength. This weight is the trust value expressing the likelihood that the trustee will successfully execute the task.

After executing the task, the trustee modifies the weight. If it completes the task successfully, it increases the weight using equation (1). If it fails, it increases the weight using equation (2).

$$w = \text{Min}\big(1, w + \alpha(1 - w)\big) \quad (1)$$

$$w = \text{Max}\big(0, w - \beta(1 - w)\big) \quad (2)$$

α, β are positive factors controlling the rate of the increase and decrease, respectively.

The trustee decides to perform a task by comparing the weight of the connection to a predefined Threshold $\in [0,1]$. In particular, it proceeds with the task's execution only in the case the weight is greater than or equal to the Threshold value.

## 2.4. The CA algorithm for dynamic trustee profiles

For the experiments reported in this paper, we employed an updated version of CA algorithm, able of dealing with the constant change of the trustee's abilities to perform tasks over time (dynamic trustee profiles). We provide algorithm's description and pseudocode as follows.

When a trustor detects a new task, it broadcasts a request message containing task information. Upon receiving a request message, every candidate trustee saves it in a list and establishes a new connection (if one does not already exist) to the trustor agent who requested the task. At each timestep, a trustee attempts to execute the task with the maximum connection weight, if the task is still available and the weight does not exceed a predetermined Threshold. If the execution is successful, the trustee increases the connection's weight; otherwise, it decreases it.

CA algorithm handles dynamic trustee profiles as follows. When a connection's weight falls below the Threshold value, the trustee interprets this as an indication of its inability to successfully execute the task and will not try it again to save resources. If the trustee's performance on an easy task is good enough indicating that the agent has probably learned how to execute difficult tasks of the same category, at which it has previously failed, it increases the connection weight of those difficult tasks to the Threshold value, giving itself a chance to try them again in the future.

---

ALGORITHM 1: CA for dynamic trustee profiles

1: **while** True **do**
    # broadcast a request message when a new task is perceived
2:  when perceived a new $task = (c, r)$
3:   broadcast message $m = (request, i, task)$
    #receive/store a request message and initialize a new connection
4:  when received a new message $m = (request, j, task)$
5:   add $m$ in list $M$
6:   **if** $\nexists$ connection $co = (i, j, \_, task)$ **then**
7:    **if** $\exists$ connection $co' = (i, \sim j, \_, task)$ **then**
8:     create $co = (i, j, avg, task)$, s.t. $avg = \frac{\sum_{\forall (i, \sim j, w, task) \in \{(i, \sim j, \_, task)\}} w}{|\{(i, \sim j, \_, task)\}|}$
9:    **else** create $co = (i, j, 0.5, task)$
10:   **end if**
11:  **end if**
    #select a task to perform
12: select message $m = (request, j, task)$ in $M$, s.t.
$\exists\ co = (i, j, w, task) \in \{(i, k, w', task)\}, \forall w': w \geq w'$
13: **if** $state\_visibility(task) = false$ or $state\_done(task) = false$ **then**
14:  **if** $canAccessAndUndertake(task) = true$ **then**
15:   **if** $w \geq Threshold$ **then**
16:    $(result, performance) \leftarrow performTask(task)$
17:   **end if**
18:  **end if**
19: **end if**
20: delete $m$ from $M$
    #update connections' weights
21: **if** $result = success$ **then**
22:  strengthen $co = (i, j, w, task)$ by using equation (1)
23: **else**
24:  weaken $co = (i, j, w, task)$ by using equation (2)

25:  **end if**
26:  $\forall\ co' = (i, j, w', task')$, s.t. $w' < Threshold$, $task' = (c, r')$, $r' > r$
27:  **if** $performance \geq minSuccessfulPerformance(task')$ **then**
28:    $w' \leftarrow Threshold$
29:  **end if**
30: **end while**

## 3. Experimental setup and methodology
### 3.1. The testbed

For the simulation experiments, we have implemented a testbed based on the one described in [5]. In this section, we describe its main characteristics and specifications.

The testbed's environment is populated with agents who provide services (referred to as providers or trustees) and agents who use these services (referred to as consumers or trustors). For the sake of simplicity, all providers deliver the same service. The agents are distributed at random in a spherical world with a radius of 1.0. Radius of operation ($r_0$) represents the agent's ability to interact with other agents. Each agent has acquaintances, i.e. other agents within its radius of operation.

The performance of the providers varies and determines the utility gain (UG) that a consumer gains from each interaction, which is calculated as follows. There are four kinds of providers: good, ordinary, bad, and intermittent. Except for the intermittent, each provider has a mean level of performance $\mu_P$, and its actual performance is normally distributed around this mean. Table 1 shows the values of $\mu_P$ and the related standard deviation $\sigma_P$, of all kinds of providers. Intermittent providers perform randomly in the range [PL_BAD, PL_GOOD]. In providers, the radius of operation is also the normal operational range within which the providers can deliver a service with no loss of quality. If a consumer is located outside that range, the provider's quality of the service degrades linearly in proportion to the distance between the provider and the consumer, but the final calculated performance always remain in [-10, +10], and is equal to the utility gained by the consumer as a result of the interaction.

In the testbed, simulations are run in rounds. As in real life, a consumer does not require the service in every round. When a consumer is created, the probability it requires the service (*activity level a*) is chosen at random. There is no other factor limiting the number of agents which can participate in a round. If a consumer needs the service in a round, the service will always be requested in that round. The time value for any event is the round number.

Each consumer falls into one of three categories: a) consumer agents without a trust model, b) consumers using FIRE, and c) consumers using CA. If a consumer agent needs to use the service during a round, it locates all nearby provider agents. Consumers without a trust model choose a nearby provider at random. FIRE consumer agents choose a provider using the four steps process described in [5]. After selecting a provider, both types of consumers, those without a trust model and those using FIRE, use the service and gain some utility. The FIRE consumers rate the service with a rating value equal to the UG they received. Then, the rating is recorded for future trust assessments. The provider is also informed about the rating and may keep it as evidence of its performance for future interactions.

On the other hand, if a CA consumer needs to use the service, it does not choose a provider. Instead, it sends a request message to all its nearby providers, defining the required quality of the service as the task's requirement. Table 2 shows the five performance levels used as the possible qualities of the service. The CA consumer first sends a message requesting the service at the highest quality (PERFECT). After a sufficient period of time (WT), all CA consumers who have not been served, broadcast a new message requesting the service at the next lowest performance level (GOOD). This process is repeated for as long as there are consumers who have not been provided with the service and the requested performance has not reached the lowest level. When a provider receives a request message, it stores it in a local list and runs CA algorithm. WT is a testbed parameter, expressing the maximum time required for all requested services of one round to be provided.

Agents can enter and exit an open MAS at any time. This is simulated by replacing a number of randomly chosen agents with new ones. The number of agents added and removed after each round varies, but they must not exceed a certain percentage of the total population. $p_{CPC}$ and $p_{PPC}$ denote the population change limits for the consumer and provider population, respectively. The characteristics of the newcomer agents are defined at random, but the proportions of different provider profiles and different consumer groups are preserved.

Changing the location of an agent affects both its own situation and its relationship with other agents. Polar coordinates $(r, \varphi, \theta)$ are used to specify the location in the spherical world. To change location, amounts of angular changes $\Delta\varphi$ and $\Delta\theta$ are added to $\varphi$ and $\theta$, respectively. $\Delta\varphi$ and $\Delta\theta$ are chosen at random in $[-\Delta\phi, +\Delta\phi]$. A consumer and a provider change their location in a round with a probability denoted by $p_{CLC}$ and $p_{PLC}$, respectively.

A provider's performance $\mu$ can be changed by an amount $\Delta\mu$ randomly chosen in $[-M, +M]$, with a probability of $p_{\mu C}$ in each round. Furthermore, with a probability of $p_{ProfileSwitch}$ a provider can change to a different profile after each round.

**Table 1 Profiles of provider agents (performance constants defined in Table 2)**

| Profile | Range of μp | σp |
| --- | --- | --- |
| Good | [PL_GOOD, PL_PERFECT] | 1.0 |
| Ordinary | [PL_OK, PL_GOOD] | 2.0 |
| Bad | [PL_WORST, PL_OK] | 2.0 |

**Table 2 Performance level constants**

| Performance level | Utility gained |
| --- | --- |
| PL_PERFECT | 10 |
| PL_GOOD | 5 |
| PL_OK | 0 |
| PL_BAD | -5 |
| PL_WORST | -10 |

### 3.2. Experimental methodology

In our experiments, we compare the performance of the following three consumer groups:

- NoTrustModel: consumer agents do not use a trust model. Instead, they select a nearby provider agent at random.
- FIRE: consumer agents use FIRE trust model.
- CA: consumers use CA trust model.

To do so, we conduct a number of independent simulation runs (NISR) for each consumer group, in order to obtain more accurate results and to avoid random noise. NISR varies in each experiment, in order to obtain statistically significant results, and its value per experiment is presented in Table 3.

The utility gained by each agent during simulations reflects the trust model's performance in identifying trustworthy provider agents. Thus, during each simulation run, the testbed records the utility gain (UG) of each consumer interaction as well as the trust model used.

After all simulation runs are completed, we compute the average UG for each interaction for each consumer group. Then, we use the two-sample t-test for means comparison [3] with a confidence level of 95% to compare the average UG of each consumer group to one another.

Each experiment's results are presented in a two-axis graph; the left axis shows the UG means of consumer groups in each interaction, and the right axis shows the performance rankings produced by the UG means comparison using the t-test. We use the group's name prefixed by 'R' to denote the plot for the performance ranking of a group. A higher-ranking group outperforms a lower-ranking group, while groups of equal rank perform insignificantly differently. For example, in Fig. 1, at the $115^{th}$ interaction (on the x-axis), agents in group CA obtain an average UG of 6.16 (reading on the left y-axis) and according to the t-test ranking, the rank of CA (as shown by the plot R.CA) is 3 (reading on the right y-axis).

We use a "typical" provider population, as defined in [5], in all the experiments, with half beneficial providers (producing positive UG) and half harmful providers (producing negative UG, including intermittent providers).

The same values for the experimental variables are used as in [5], shown in Table 4, while Table 5 and Table 6 show the default FIRE and CA parameters used, respectively.

**Table 3 Number of independent simulation runs (NISR) per experiment**

| Experiment | NISR |
|---|---|
| 1, 2, 6, 8, 10, 11 | 30 |
| 3-5, 7 | 10 |
| 9 | 12 |

**Table 4 Experimental variables**

| Simulation variable | Symbol | Value |
|---|---|---|
| Number of simulation rounds | $N$ | |
| - Default |  | 500 |
| - Experiments: 7-9 |  | 1000 |
| Total number of provider agents: | $N_P$ | 100 |
| Good providers | $N_{GP}$ | 10 |
| Ordinary providers | $N_{PO}$ | 40 |
| Intermittent providers | $N_{PI}$ | 5 |
| Bad providers | $N_{PB}$ | 45 |
| Total number of consumer agents | $N_C$ | 500 |
| Range of consumer activity level | α | [0.25, 1.00] |
| Waiting Time | WT | 1000 msec |

**Table 5 FIRE's default parameters**

| Parameters | Symbol | Value |
|---|---|---|
| Local rating history size | $H$ | 10 |
| IT recency scaling factor | $\lambda$ | -(5/ln(0.5)) |
| Branching factor | $n_{BF}$ | 2 |
| Referral length threshold | $n_{RL}$ | 5 |
| Component coefficients |  |  |
| Interaction trust | $W_I$ | 2.0 |
| Role-based trust | $W_R$ | 2.0 |
| Witness reputation | $W_W$ | 1.0 |
| Certified reputation | $W_C$ | 0.5 |
| Reliability function parameters |  |  |
| Interaction trust | $\gamma_I$ | -ln(0.5) |
| Role-based trust | $\gamma_R$ | -ln(0.5) |
| Witness reputation | $\gamma_W$ | -ln(0.5) |
| Certified reputation | $\gamma_C$ | -ln(0.5) |

**Table 6 CA's default parameters**

| Parameters | Symbol | Value |
|---|---|---|
| Threshold | $Threshold$ | 0.5 |
| Positive factor controlling the rate of the increase in strengthening of a connection | $\alpha$ | 0.1 |
| Positive factor controlling the rate of the decrease in weakening of a connection | $\beta$ | 0.1 |

## 4. Simulation results
### 4.1. CA's vs. FIRE's performance in the static setting

The first experiment serves as a baseline experiment and its purpose is to compare the performance of CA to that of FIRE and a group of agents with no trust model in a static setting with no dynamic factors in effect.

We specifically investigate whether CA helps the consumers achieve higher utility gain than the other two groups. Fig. 1 shows that the NoTrustModel group performs the worst in all interactions. CA and FIRE, on the other hand, both help consumers in gaining far greater UG, demonstrating that both models are capable of learning how to pick profitable provider agents. The chart, as well as the t-test ranking, shows that FIRE can learn about the providers faster than CA in the first 38 interactions, since it achieves a greater UG than CA. Yet, after interaction 38, CA consistently outperforms FIRE with a rather stable performance. On the contrary, FIRE's performance is steadily improving tending to surpass CA's performance after interaction 480. Yet, larger-scale experiments (e.g. N=1000) are required to determine whether FIRE outperforms CA after interaction 480.

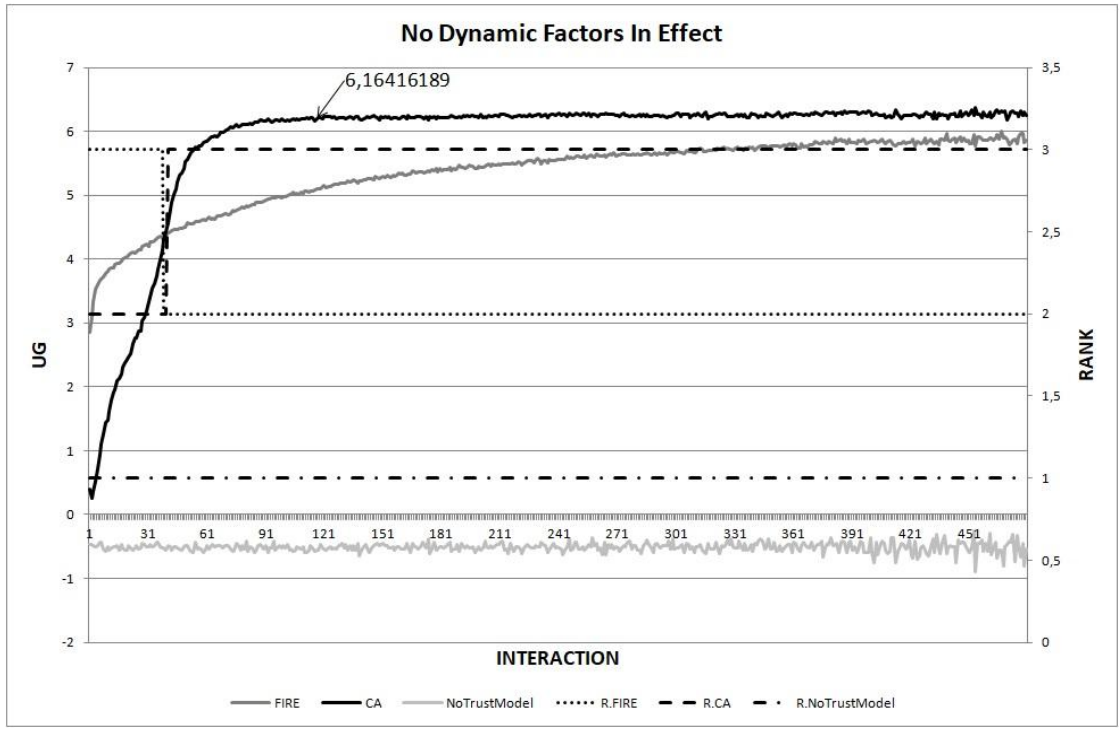

Fig. 1 Experiment 1: CA's vs. FIRE's performance in the static setting

### 4.2. CA's vs. FIRE's performance in provider population changes

In this section, we compare the performance of CA and FIRE, when the population of providers gradually changes up to 10%, by conducting the following three experiments.

Experiment 2. The provider population changes at maximum 2% in every round ($p_{PPC} = 0.02$).

Experiment 3. The provider population changes at maximum 5% in every round ($p_{PPC} = 0.05$).

Experiment 4. The provider population changes at maximum 10% in every round ($p_{PPC} = 0.10$).

The results of these experiments are shown in figures Fig. 2, Fig. 3 and Fig. 4 . Overall, both models continue to help the consumers in obtaining positive UG, whereas the NoTrustModel group consistently performs the worst. Nevertheless, the provider population change has a negative impact on both models' performance, which is worse than that in the static setting (Fig. 1).

More specifically, in Experiment 2 (Fig. 2), both the chart and the t-test ranking show that FIRE is superior in the first 41 interactions, while CA outperforms FIRE with a stable performance after interaction 41. FIRE, on the other hand, is constantly improving to reach CA's performance.

Experiment 3 (Fig. 3) and experiment 4 (Fig. 4) show that when the provider population changes more dramatically, CA performs worse than FIRE. Figures Fig. 5 and Fig. 6 show more clearly, that FIRE is more resilient to this type of environmental change. As the provider population change rate increases from 2% to 10%, FIRE's performance drops by only one UG unit, while CA's performance drops by four. This is because CA model is based on the providers' knowledge of their capabilities and newcomer provider agents must learn from scratch.

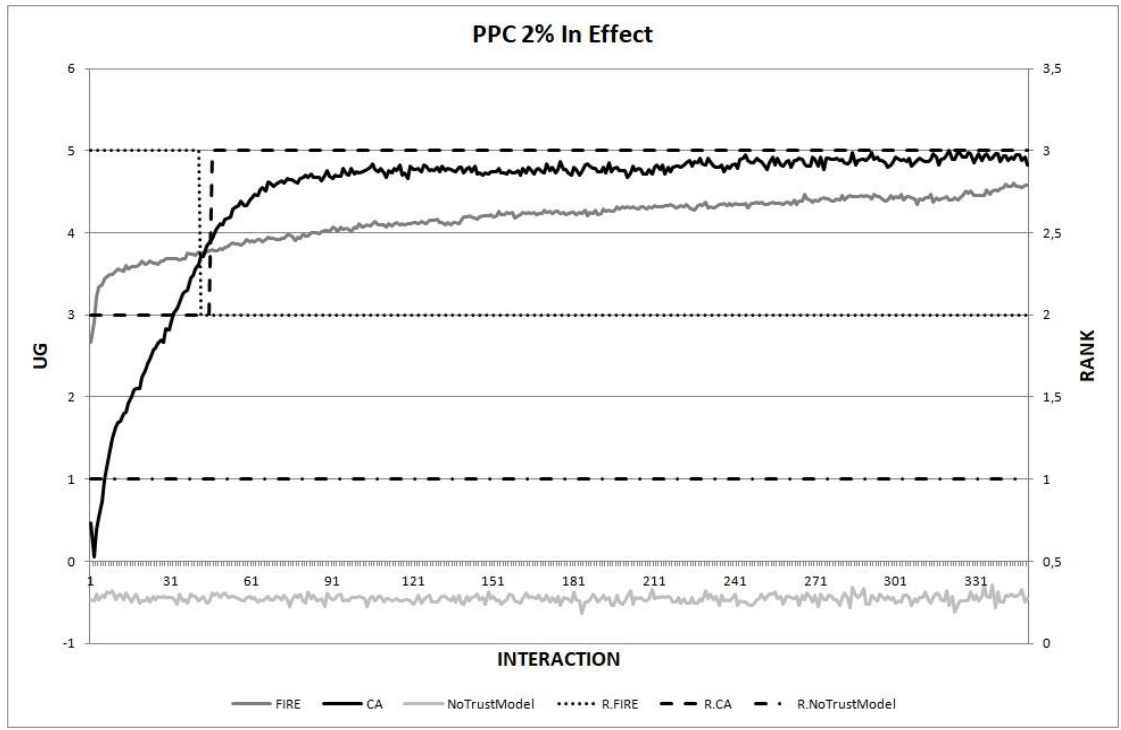

Fig. 2 Experiment 2: Provider population change $p_{PPC} = 2\%$

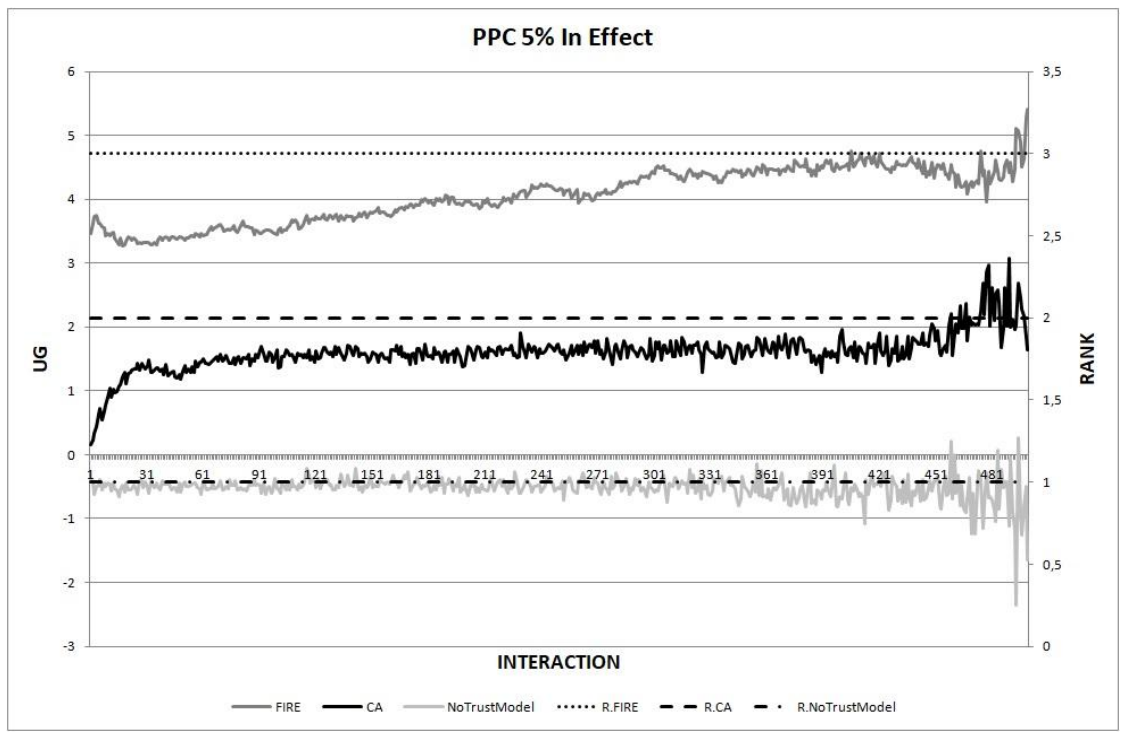

Fig. 3 Experiment 3: Provider population change $p_{PPC} = 5\%$

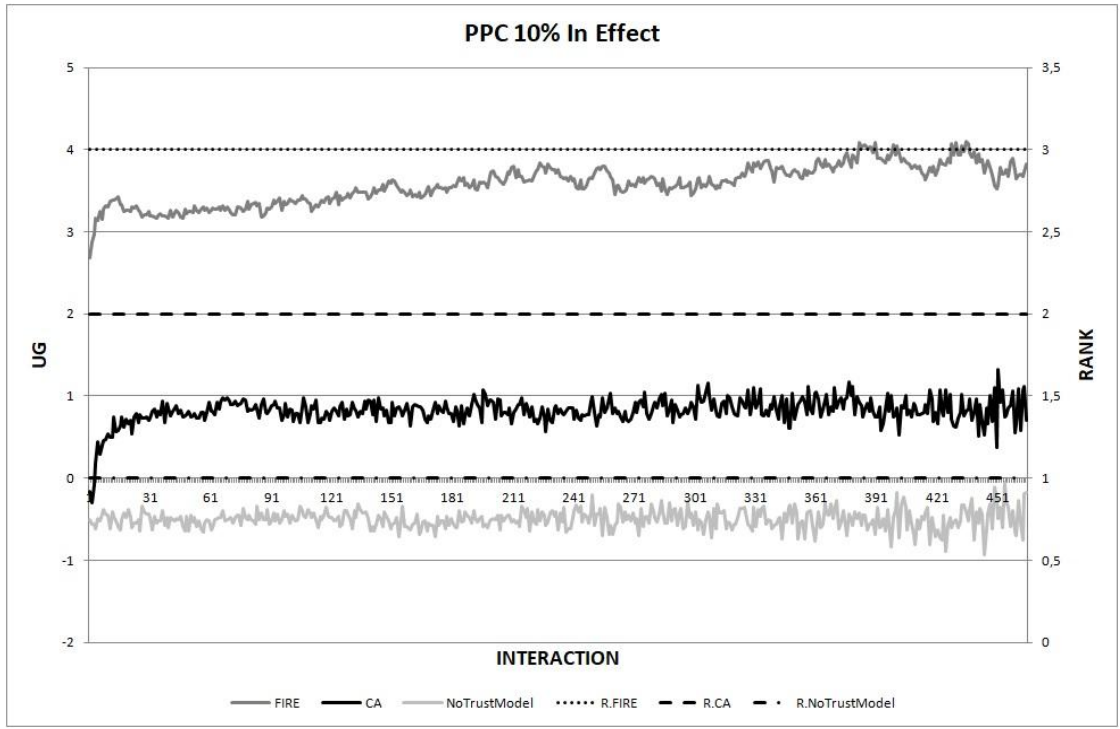

Fig. 4 Experiment 4: Provider population change $p_{PPC} = 10\%$

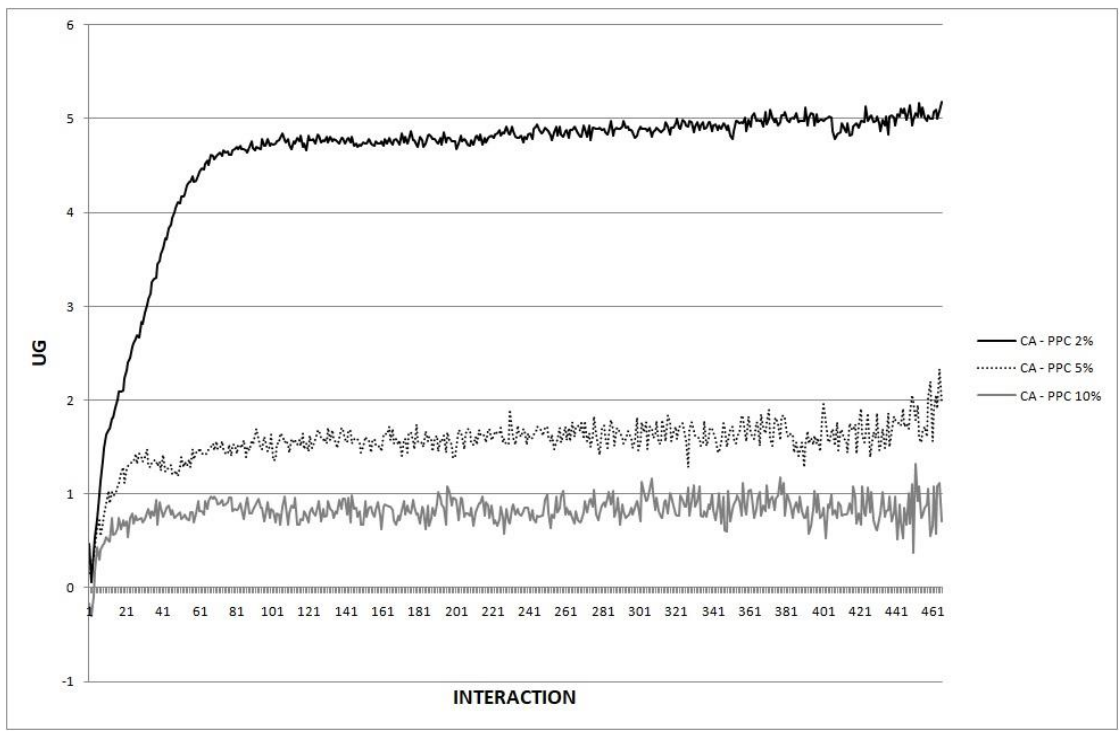

Fig. 5 As the provider population change rate increases from 2% to 10% CA's performance drops 4 UG units.

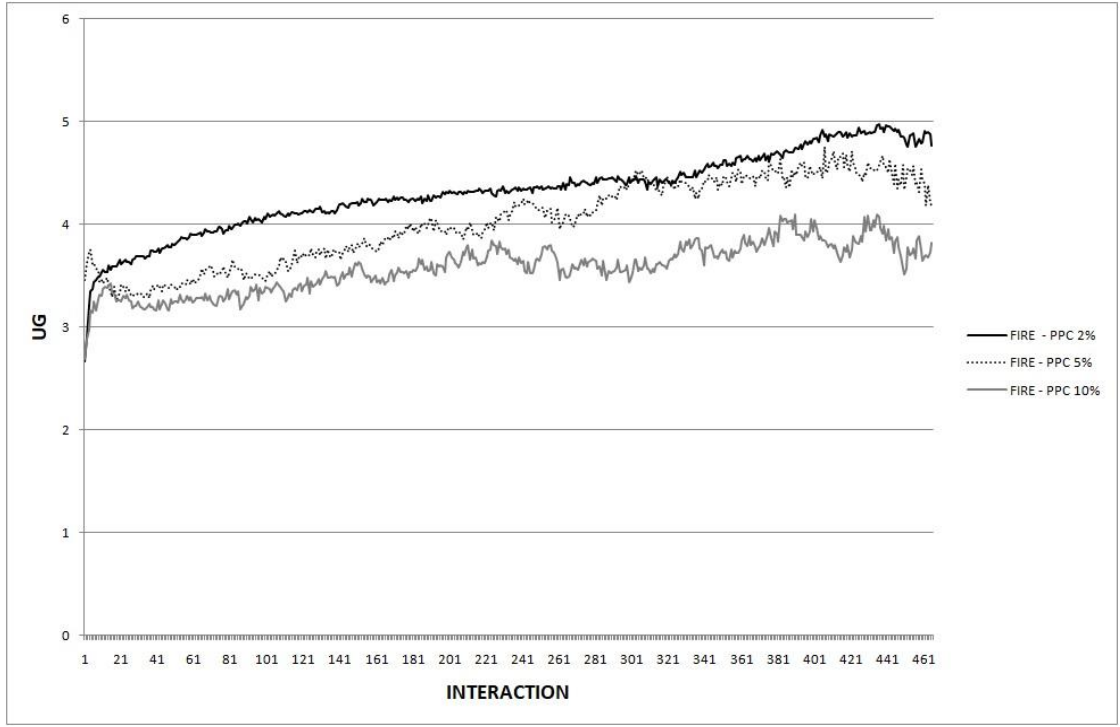

Fig. 6 As the provider population change rate increases from 2% to 10% FIRE's performance drops only by 1 UG unit

### 4.3. CA's vs. FIRE's performance in consumer population changes

In this section, we compare the performance of CA and FIRE, when the population of consumers changes up to 10% by conducting the following three experiments.

Experiment 5. The consumer population changes at maximum 2% in every round ($p_{CPC} = 0.02$).

Experiment 6. The consumer population changes at maximum 5% in every round ($p_{CPC} = 0.05$).

Experiment 7. The consumer population changes at maximum 10% in every round ($p_{CPC} = 0.10$).

Figures Fig. 7, Fig. 8, Fig. 9 show the experiments' results. Overall, both models, CA and FIRE continue to help the consumers in obtaining positive UG, whereas the NoTrustModel group consistently performs the worst.

More specifically, in Experiment 5 (Fig. 7), CA outperforms FIRE from the first interaction with a rather steady performance at the levels of its performance in the static environment. FIRE on the other hand, performs worse than in the static setting. As the consumer population change rate increases gradually from 2% to 10% CA consistently outperforms FIRE. This is explained by the fact that FIRE, unlike CA, is dependent on the consumer population for Witness Reputation, and thus performs worse than CA. Figure Fig. 10 shows that CA is not only resilient to this kind of change, but also that its performance improves with the increase of provider population change rate.

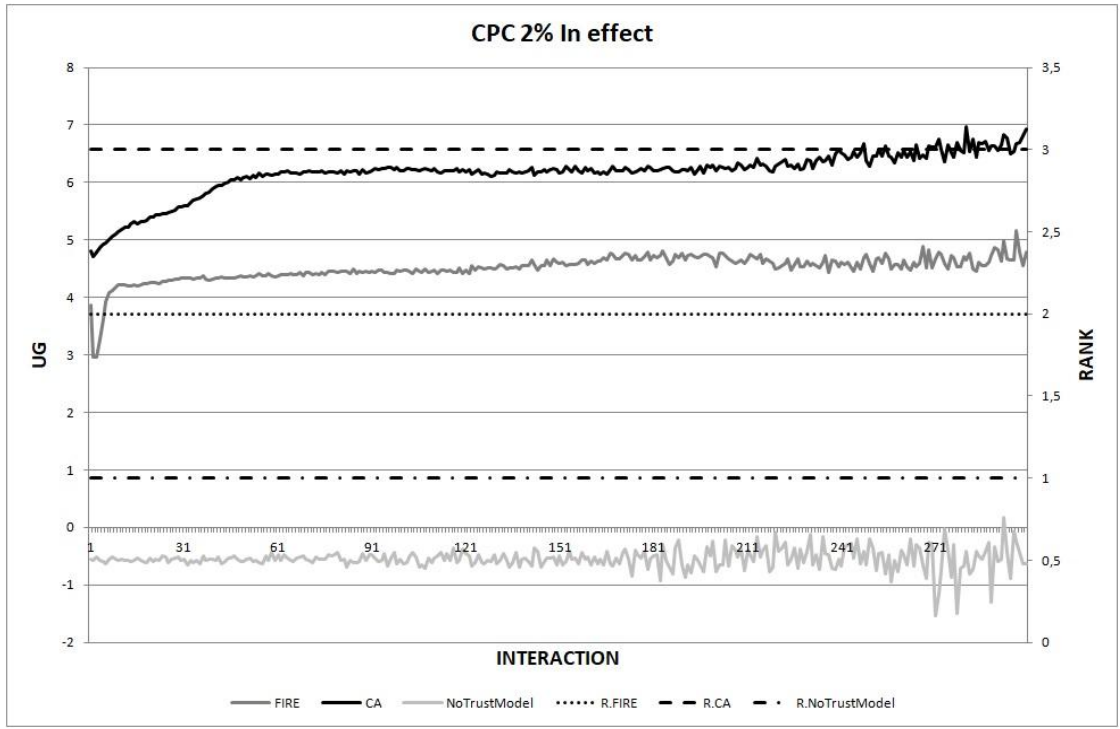

Fig. 7 Experiment 5: Consumer population change $p_{CPC} = 2\%$

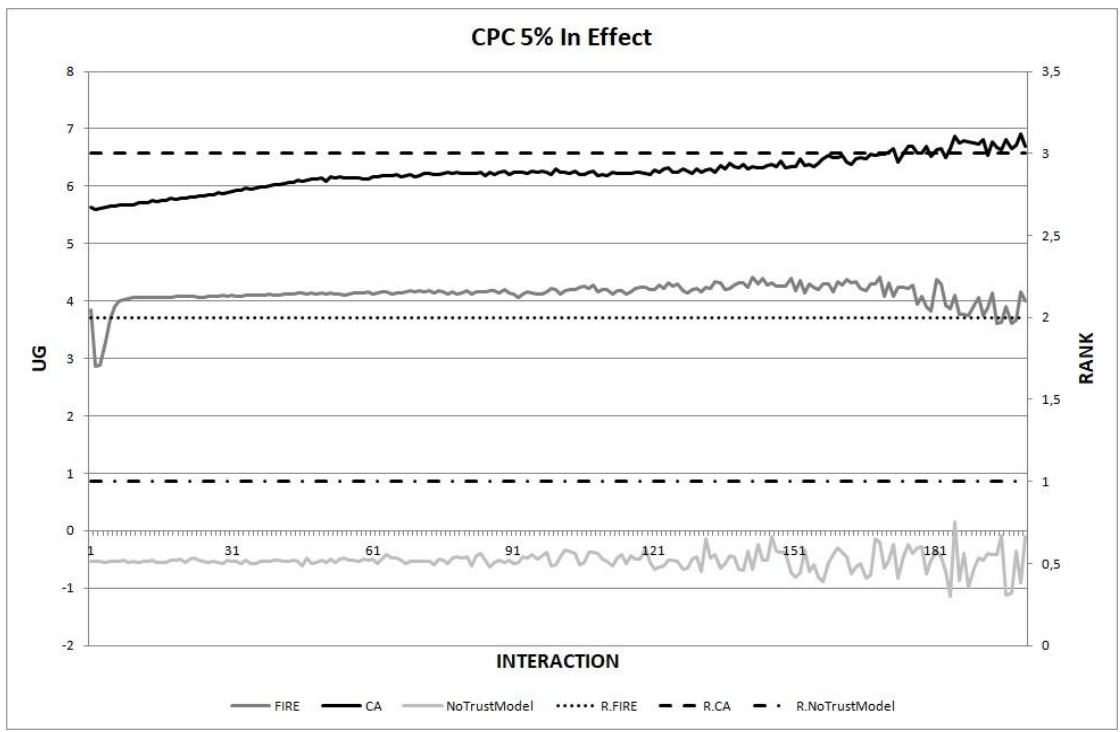

Fig. 8 Experiment 6: Consumer population change $p_{CPC} = 5\%$

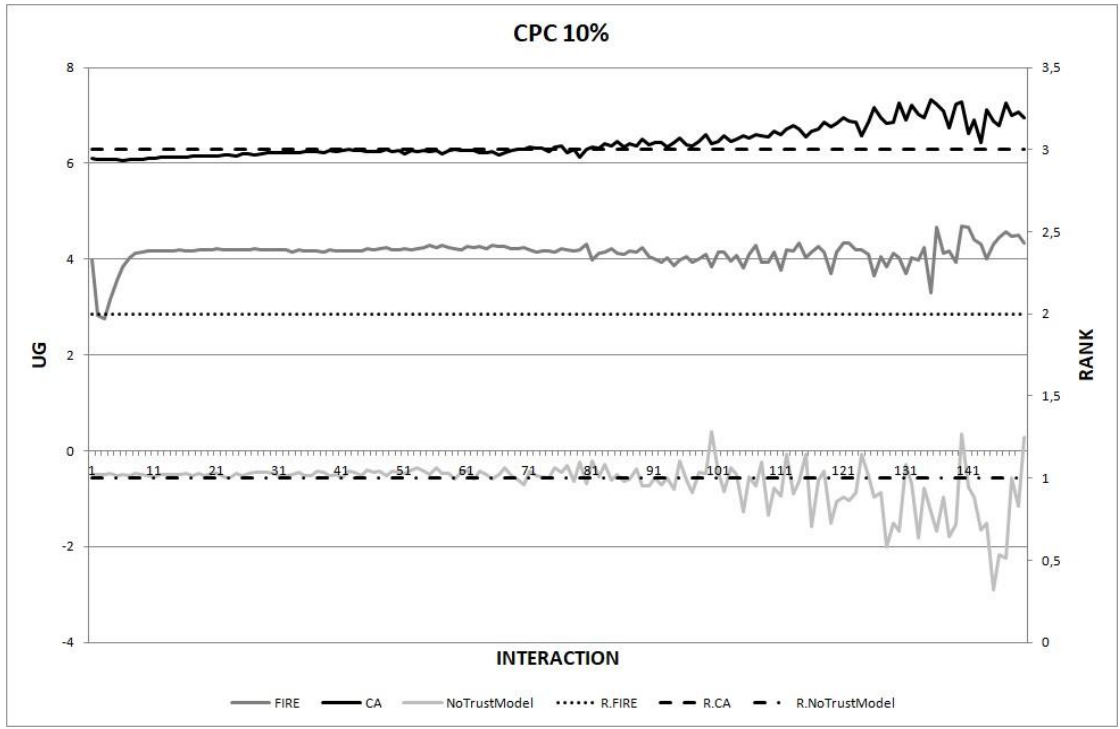

**Fig. 9** Experiment 7: Consumer population change $p_{CPC} = 10\%$

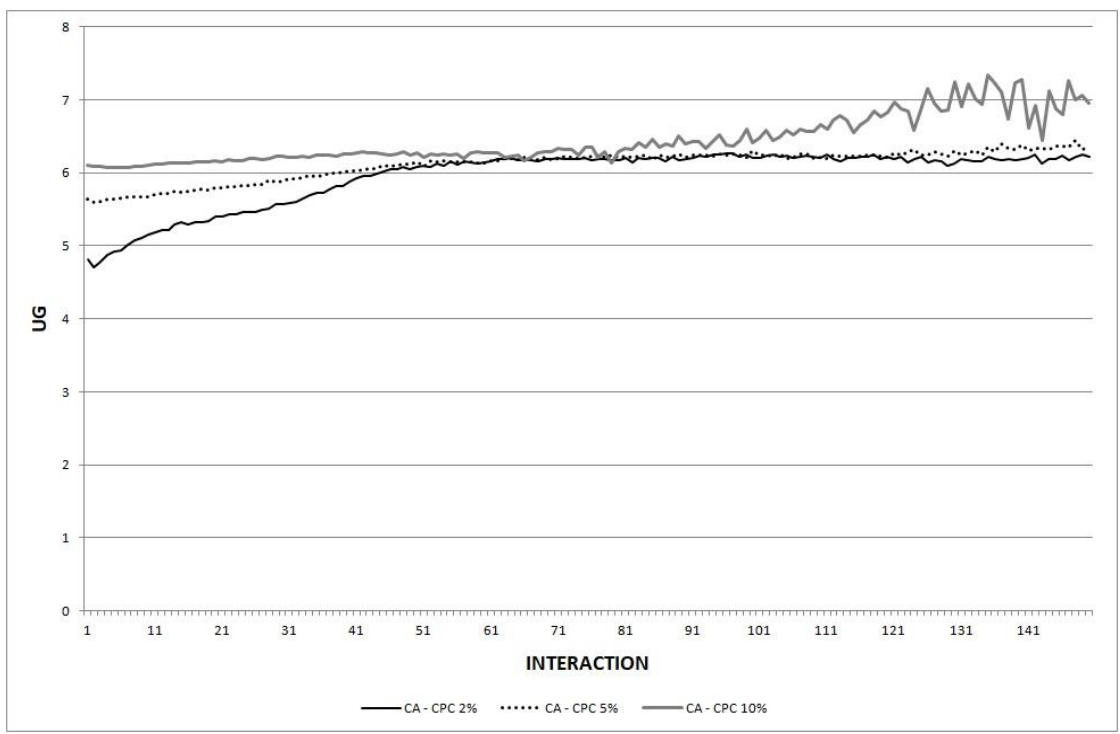

**Fig. 10** CA's performance improves as the consumer population change rate increases from 2% to 10%

### 4.4. The effect of concurrently changing consumer and provider populations

In this section, we report the results of two experiments, in which we test the resilience of both models when the consumer and provider populations change at the same time. The experiments are as follows.

Experiment 8. The provider population changes at maximum 2% in every round ( $p_{PPC}$ = 0.02) and the consumer population changes at maximum 5% in every round ($p_{CPC}$ = 0.05).

Experiment 9. The provider population changes at maximum 10% in every round ( $p_{PPC}$ = 0.10) and the consumer population changes at maximum 10% in every round ($p_{CPC}$ = 0.10).

Experiment 9 (Fig. 12) shows that while CA is better in the first few interactions, FIRE eventually prevails. This is in contrast to the results of Experiment 8 (Fig. 11), in which CA outperforms throughout all interactions. As the population change becomes more severe, FIRE's performance is slightly affected, whereas CA's performance drops significantly, by 3 UG units.

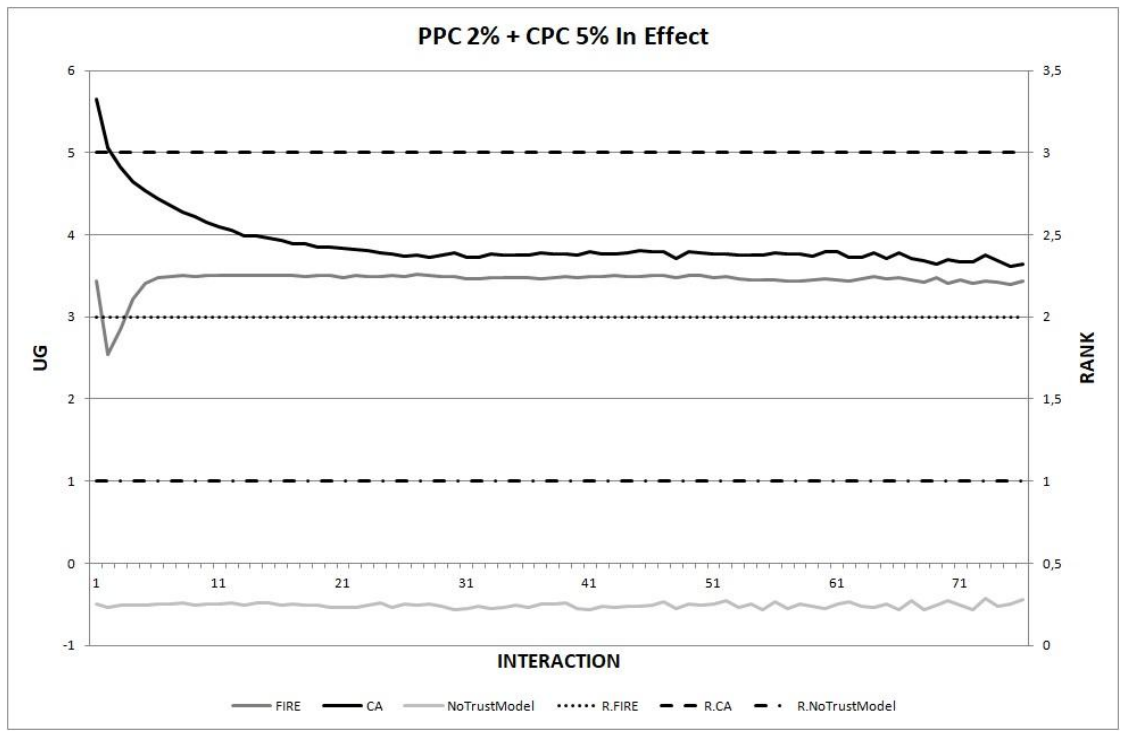

**Fig. 11 Experiment 8: $p_{PPC} = 2\%$ and $p_{CPC} = 5\%$. CA outperforms in all interactions.**

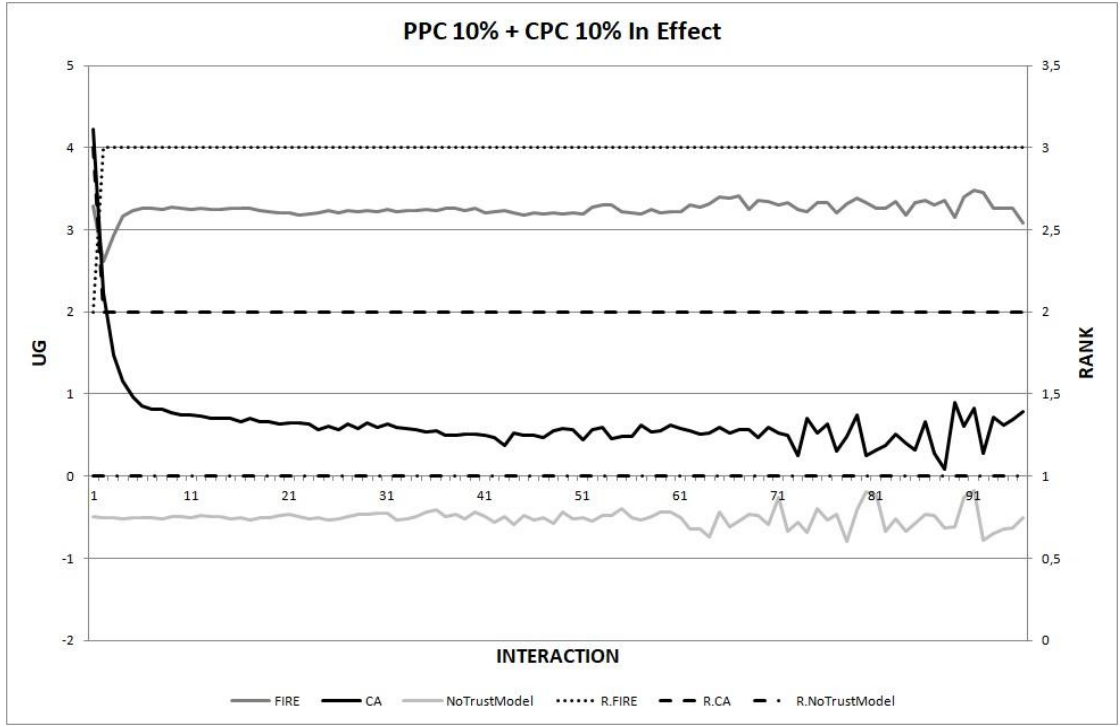

Fig. 12 Experiment 9: $p_{PPC} = 10\%$ and $p_{CPC} = 10\%$. CA performs better in the first interactions, but FIRE eventually outperforms.

### 4.5. CA's vs. FIRE's performance in dynamic trustees profiles

In this section, we report the results of two experiments, in which we test the performance of CA and FIRE, when the trustees change their abilities to perform tasks over time. The experiments we conducted are as follows.

Experiment 10. A provider may alter its average level of performance at maximum 1.0 UG unit with a probability of 10% in every round ($p_{\mu C} = 0.10, M = 1.0$).

Experiment 11. A provider may switch into a different (performance) profile with a probability of 2% in every round ($p_{ProfileSwitch} = 0.02$).

In Experiment 10 (Fig. 13) providers change their performance levels. CA's performance begins with almost zero UG and rapidly increases until interaction 60, when it stabilizes at an average UG of around 6.6. In contrast, FIRE's performance begins higher (UG = 2.95) and gradually improves to surpass CA'S performance after interaction 285.

Experiment 11 (Fig. 14), in which providers change performance profiles, has the most negative effect on the performance of both models. FIRE outperforms throughout all interactions, with an average UG of 3.15. On the other hand, CA has a worse performance with an average UG of 2.64.

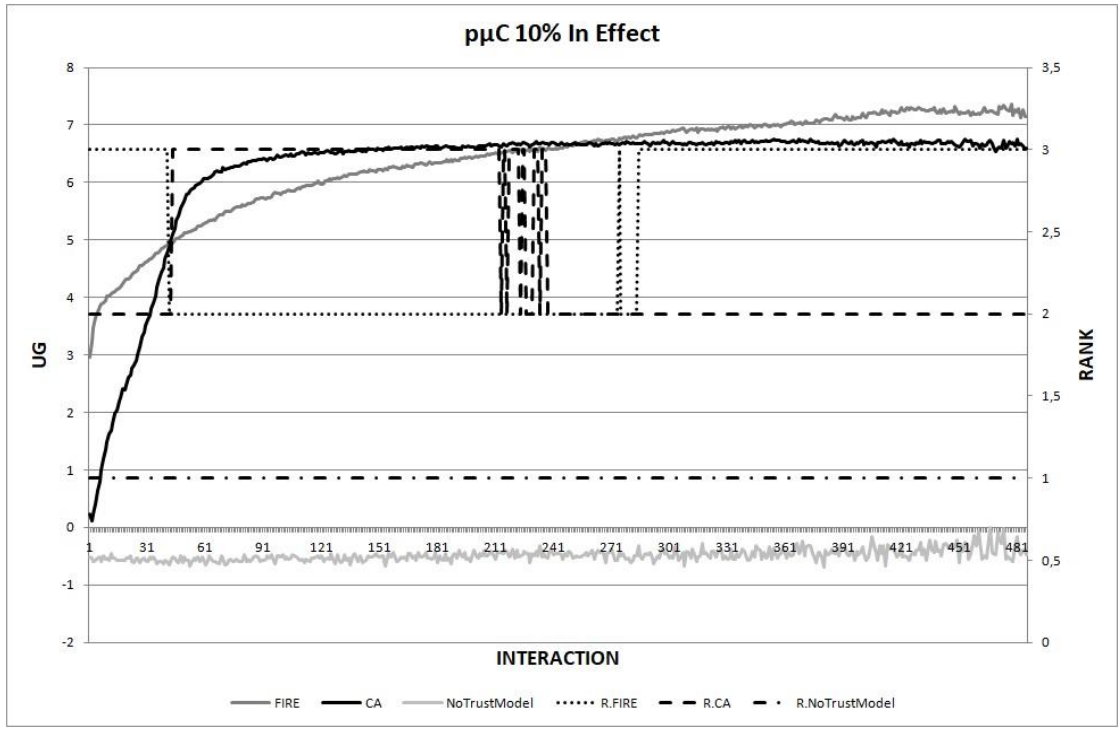

Fig. 13 Experiment 10: Providers change their performance ($p_{\mu C} = 10\%, M = 1.0$)

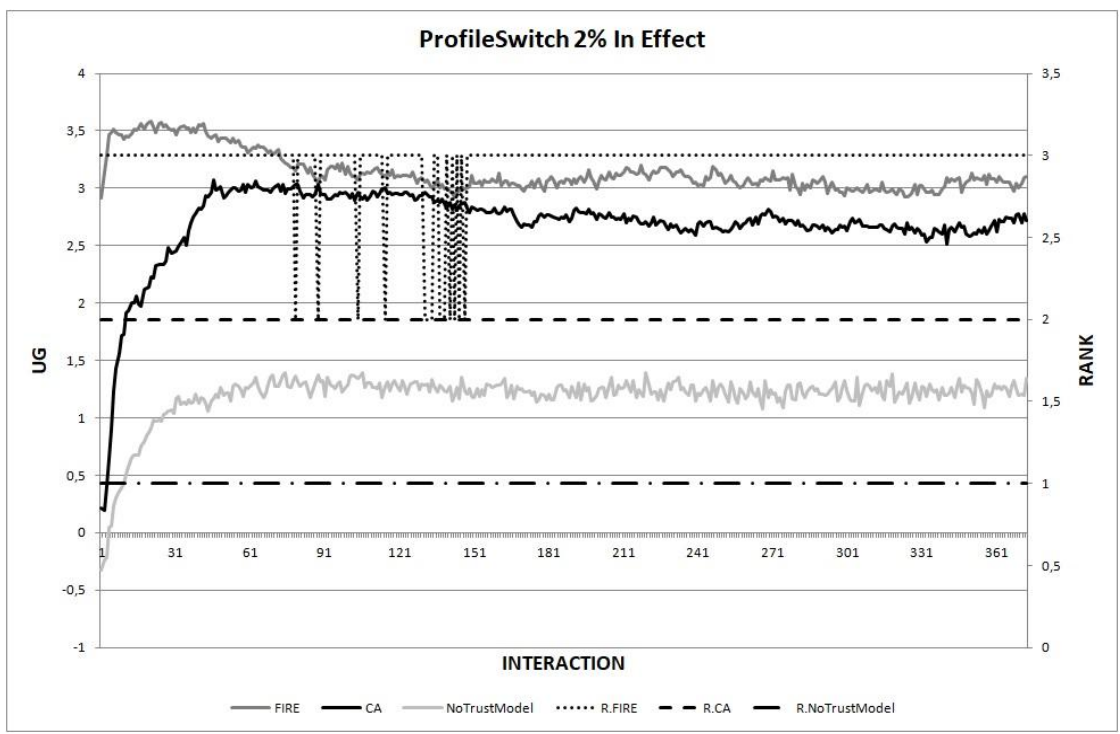

Fig. 14 Experiment 11: Providers switch their profiles ($p_{ProfileSwitch} = 2\%$)

## 5. Discussion

According to the results presented in Section 4, CA model has a clear advantage in conditions of change in the population of consumers, while FIRE is advantageously placed to respond to the perturbations caused by the provider's population change. In this section, we discuss the reasons for these findings.

FIRE is a trust and reputation model that integrates four different sources of trust information (Section 2.2). Interaction trust, in which a provider's trustworthiness is mainly determined based on the evaluator's previous interactions with the target agent, is the most reliable form of trust, since it reflects the evaluator's satisfaction. However, in the case that the evaluator has no previous interactions, FIRE cannot use the Interaction Trust module and relies on the other three modules, and mainly on Witness Reputation, in which the target's trustworthiness is estimated based on the opinions of other agents (witnesses) that have interacted with the target. However, when witnesses leave the system, they take with them their opinions and Witness Reputation module cannot work. In contrast, CA does not rely on the opinions of other consumer-witnesses, and this is the reason for its resilience in changes in the population of consumers, even in large ones.

CA model is based on the providers' knowledge of their capabilities, which is stored in the form of weights of the connections they create. A provider's longer stay in the system results in more interactions and thus, a great number of connections with weights that reflect its true ability to provide the service. CA's performance improves as the number of providers who are aware of their true capabilities increases. However, when provider replacements occur frequently, CA's performance is significantly hampered by bad newcomers who provide bad quality services, as they are unfamiliar with their capabilities and must learn from scratch.

It is not totally surprising, then, that while the CA model has been conceived with the motivation of allowing trustees to play a more significant role in trust mechanisms, it is in environments where the trustee population is relatively stable that this initial motivation is, eventually, realized to the full potential; accordingly, when the trustee population is volatile, a model that places emphasis on trustors is more resilient.

6.  **Conclusions and future work**

Existing trust and reputation models still suffer from a number of serious unresolved issues, including the inability to handle continuously changing behaviors and agents' continuous entering – exiting of the system. CA, is a biologically inspired, decentralized computational trust model, suitable for dynamic open MAS. The distinguishing feature that gives rise to its advantages is that it works from the trustee's perspective, i.e. the choice is devoid from the trustor and the trustee decides whether it is capable of providing a service.

We ran a series of simulations to compare CA model to FIRE, a well-established, decentralized trust and reputation model for open MAS under several dynamically changing environmental factors, such as the continuous entering and leaving of agents, and the changing of agents' behavior and profiles.

Our main results can be summarized as follows. CA has a stable performance in the static setting and it can still locate the profitable providers in an environment with a variety of dynamic factors in effect, maintaining a high UG for the consumer population. FIRE is more resilient to changes in provider population than CA, which does not respond well to this type of environmental change, because it is based on the providers' knowledge of their capabilities and newcomer providers must learn from scratch. CA, on the other hand, is unaffected by the consumer population change. This is due to the fact that CA, unlike FIRE, does not rely on consumers for Witness Reputation and thus performs better. The environmental change with the most dramatic effect in both models' performances is when the providers switch performance profiles, where FIRE is more resilient.

Reasoning about openness and particularly agents' evolution (Section 2.1), not only requires considering possible changes to one's capabilities, but also reasoning about changes to one's own type. A direction for future work is to investigate how CA algorithm could be modified to handle switches of performance profiles more effectively. Neither CA nor FIRE outperforms with any combination of varying degrees of dynamic factors in effect. Thus, a natural rich area of future research is to investigate how the consumers can detect the dynamic factors present in their environment and then decide which trust model to employ in order to maximize utility gain.